\documentclass{llncs}

\newcommand{\proofsketch}{{\scshape\noindent Proof (sketch).\quad}}
\renewcommand\equiv{:=}

\usepackage{a4wide}
\usepackage{amssymb, amsmath}
\usepackage[utf8]{inputenc}
\usepackage{epsfig}
\usepackage{color}
\usepackage[all]{xy}

\newtheorem{lemm}{Lemma}
\newtheorem{theo}[lemm]{Theorem}
\newtheorem{corr}[lemm]{Corollary}
\newtheorem{prop}{Proposition}
\newtheorem{conj}{Conjecture}

\newcommand*\pop{\mbox{\textsf{pop}}}
\newcommand*\push{\mbox{\textsf{push}}}
\newcommand*\zcopy{\mbox{\textsf{copy}}}

\newcommand\dom{\mbox{\textsf{dom}}}
\newcommand\dec{\mbox{\textsf{dec}}}
\newcommand\inc{\mbox{\textsf{inc}}}
\newcommand\tail{\mbox{\textsf{tail}}}
\newcommand*\id{\mbox{\textsf{id}}}

\newcommand*{\vers}[1]{\stackrel{#1}{\longrightarrow}} 
\newcommand*{\chemin}[1]{\xrightarrow{#1}} 
\renewcommand*{\graph}[1]{\mathcal H_{#1}} 
\newcommand*\g[1]{\mathcal G_{#1}} 

\newcommand*\treegraph{\textnormal{Treegraph}}
\newcommand*\MTh{\textnormal{MTh}}

\newcommand*\wII{{\omega\Uparrow}} 
\newcommand*\wIIn{{\omega\Uparrow n}}
\newcommand*{\w}{{\omega}}
\newcommand*{\ww}{{\omega^\omega}}
\newcommand*{\www}{\omega^{\omega^\omega}}
\newcommand*{\II}{\Uparrow}
\renewcommand*{\phi}{\varphi}
\newcommand*\e{{\varepsilon_0}}
\newcommand\rh{\prec} 

\title{Covering of ordinals}
\author{Laurent Braud}
\institute{\small Institut Gaspard Monge, Université Paris-Est}
\begin{document}
\maketitle

\begin{abstract}
The paper focuses on the structure of fundamental sequences of
ordinals smaller than $\e$. A first result is the construction of a
monadic second-order formula identifying a given structure, whereas
such a formula cannot exist for ordinals themselves. The structures
are precisely classified in the pushdown hierarchy. Ordinals are also
located in the hierarchy, and a direct presentation is given.
\end{abstract}

A recurrent question in computational model theory is the problem of
model checking, i.e. the way to decide whether a given formula holds
in a structure or not.  When studying infinite structures, first-order
logic only brings local properties whereas second-order logic is most
of the time undecidable, so monadic second-order logic or one of its
variants is often a balanced option. In the field of countable
ordinals, results of Büchi \cite{buchi73} and Shelah \cite{shelah}
both brought decidability of the monadic theory via different ways.
This positive outcome is tainted with the following property~: the
monadic theory of a countable ordinal only depends on a small portion
of it, called the $\omega$-tail \cite[Th. 4.9]{buchi73}. In other
words, many ordinals greater than $\ww$ share the same monadic
theories and cannot be distinguished.

Another class of structures enjoying a decidable monadic second-order
theory is the pushdown hierarchy \cite{caucal02}, which takes its
source in the Muller and Schupp characterization of transition graphs
of pushdown automata \cite{mullerschupp}. In the same way, each level
of the hierarchy has two characterizations\,: an internal by
higher-order pushdown automata \cite{regularpushdown}, and an external
presentation by graph transformations \cite{carayolwohrle}. This paper
will use the latter by the means of monadic interpretation and
treegraph operations.

The original motivation of this paper was the localization of ordinals
smaller than $\e$ in the hierarchy. Because of the above property,
ordinals themselves are not easy to manipulate with monadic
interpretations. There is therefore a need of structures as expressive
as ordinals (in terms of interpretations) but having additional
properties, such as the existence of a monadic formula precisely
identifying the structure.

A well-known object answers to this request. Each countable limit
ordinal may be defined as the limit of a so-called fundamental
sequence. For ordinals smaller than $\e$, it is easy to have a unique
definition for this sequence using the Cantor normal form. We note
$\alpha\rh\beta$ when $\alpha$ is in the fundamental sequence of
$\beta$ or $\alpha+1=\beta$.  When restricted to ordinals smaller than
$\lambda$, we call the resulting structure the covering graph of
$\lambda$. In Section \ref{covering}, we present precisely this
structure and give some of its properties. In particular, the
out-degree of its vertices is studied intensively. This eventually
yields a specific formula for each covering graph.

Section \ref{hierarchy} locates the covering graph of any ordinal
$\alpha$ smaller than $\e$ in the level $n$ of the hierarchy, where
$n$ is the largest size of the $\omega$-tower smaller than
$\alpha$. The result also applies to ordinals themselves. This was
already shown for ordinals up to $\omega^\ww$ in \cite{esikbloom}.  In
Section \ref{strict}, the result in strengthened by proving that
covering graphs are not in the lower levels; the question is still
open for ordinals. Eventually, we produce a direct presentation for
towers of $\w$ through prefix-recognizable relations of order $n$, but
involving a more technical proof.

Similar attempts of characterization of ordinals has
been made in the field of automaticity \cite{Delh04,KRS05}, but in the
other way around~: word- and tree-automatic ordinals are shown to be
respectively less than $\ww$ and $\www$.

\section{Definitions}
\label{def}

In this paper, ordinals are often considered from a graph theory point
of view. The set of vertices of $\alpha$ is the set of ordinals
smaller than $\alpha$, and the set of arcs is the relation $<$.

\subsection{Graphs}
Graphs are finite or infinite sets of labeled arcs.  A $\Sigma$-graph
is a set $G\subseteq V\times\Sigma\times V$, where $V$ (or $V_G$ if
unclear) is the \emph{support}, i.e. a finite or countably infinite
set of \emph{vertices}, and $\Sigma$ a finite set of \emph{labels}. An
element $(p,a,q)$ of $G$ is called an \emph{arc} and noted
$p\vers{a}q$.  Each label $a\in\Sigma$ is associated to a relation
$R_a = \{(p,q)\;|\;p\vers{a}q\}$ on $V$. A finite sequence of arcs
$p\vers{a_1}\dots\vers{a_n}q$ is a \emph{path} and noted
$p\chemin{a_1\dots a_n}q$.  This is extended to languages with
$p\chemin{L}q$ iff $\exists u\in L$ such that
$p\chemin{u}q$. Isomorphism between graphs is noted $\simeq$.

The \emph{monadic second-order} (MSO) logic is defined as usual; see
for instance \cite{lncs2500}. We take a set of (lowercase) first-order
variables and a set of (uppercase) second-order variables. For a given
set of labels $\Sigma$, atomic formulas are $x\in X$, $x=y$ and
$x\vers{a}y$ for all $a\in\Sigma$ and $x,y,X$ variables. Formulas are
then closed by the propositional connectives $\neg, \land$ and the
quantifier $\exists$.  Graphs are seen as relational structures over
the signature consisting of the relations $\{R_a\}_{a\in\Sigma}$.  The
set of closed monadic formulas satisfied by a graph $G$ is noted
$\MTh(G)$.

Given a binary relation $R$, the \emph{in-degree} (respectively
\emph{out-degree}) of $x$ is the cardinality of $\{y\,|\,yRx\}$
(resp. $\{y\,|\,xRy\}$). The output degree in a graph $G$ of $x\in V$
is the cardinal of $\{y\,|\,\exists a, (x,a,y)\in G\}$. The output
degree of a graph is the maximal output degree of its vertices if it
exists.

\subsection{Ordinals}

For a general introduction to ordinal theory, see
\cite{rosenstein,roitman}. An order is a well-order when each
non-empty subset has a smallest element. Ordinals are well-ordered by
the relation $\in$, and satisfy $\forall x(x\in\alpha\Rightarrow
x\subset\alpha)$. Since any well-ordered set is isomorphic to a unique
ordinal, we will often consider an ordinal up to isomorphism. 
In terms of graphs, the set of labels of an ordinal is a singleton
often noted $\Sigma = \{<\}$ and the graph respects the following
monadic properties :

\[\begin{array}{rc}
\textrm{(strict order)}&\left\{\begin{array}{c}
\forall p,q(\neg(p\vers{<}q\land q\vers{<}p))\\
\forall p,q,r((p\vers{<}q\land(q\vers{<}r)\Rightarrow p\vers{<}r)\\
\end{array}\right.\\
\textrm{(total order)}&\forall p,q(p\vers{<}q\lor q\vers{<}p\lor p=q)\\
\textrm{(well order)}&\forall X\neq\emptyset\;\exists x(x\in X\land\forall y(y\in X\Rightarrow(x\vers{<}y\lor x=y)))\\
\end{array}\]

The ordinal arithmetics define operations on ordinals such as
addition, multiplication, exponentiation.  The bound of ordinals
investigated here is $\e$, the smallest ordinal such that
$\e=\omega^\e$; therefore {\bf the declaration ``$<\e$'' is implicit
  through the rest of the paper}. To simplify the writing of towers of
$\omega$, the notation $\Uparrow$ is used to note the iteration of
exponentiation ie.  $a\Uparrow b =
\left.a^{a^{\dots^a}}\right\}{b\textrm{ times}}$. In particular,
$a\Uparrow 0=1$ is the (right) exponentiation identity.

Classic operations are not commutative in ordinal theory : for
instance $\omega+\omega^2 = \omega^2< \omega^2+\omega$. This leads to
many writings for a single ordinal.  Fortunately, all ordinals smaller
than $\e$ may uniquely be written in the Cantor normal form (CNF)
\[
\alpha = \omega^{\alpha_0}+\dots+\omega^{\alpha_k}
\]
where $\alpha_k\leq\dots\leq\alpha_0<\alpha$. An alternative we will
call \emph{reduced} Cantor normal form (RCNF) is $\alpha =
\omega^{\alpha_0}.c_0+\dots+\omega^{\alpha_k}.c_k$ where
$\alpha_k<\dots<\alpha_0<\alpha$ and $c_1,\dots,c_k$ are non-zero
integers. To express ordinals smaller than $\e$ from natural numbers
and $\omega$, the only operations needed are thus addition and
exponentiation.



\section{Covering graphs}
\label{covering}

In this section, we define the covering graph of an ordinal as the
graph of successor and fundamental sequence relations. Then, we prove
some of its important properties. One of them is the finite degree
property, which is worked out to bring a specific monadic formula for
each covering graph, thus allowing to differentiate them.

\subsection{Fundamental sequence}

The cofinality \cite{rosenstein} of any countable ordinal is
$\omega$. To each limit ordinal $\alpha$ we may associate a
$\omega$-sequence whose bound is $\alpha$. For $\alpha\leq\e$, $\alpha
= \beta + \omega^{\gamma}$ with $\beta<\alpha$, $\gamma<\alpha$ and
$\omega^\gamma$ is the last term in the CNF of $\alpha$, we define the
fundamental sequence $(\alpha[n])_{n<\omega}$ as follows\;:

\[\alpha[n] = \left\{\begin{array}{ll}
\beta + \omega^{\gamma'}.(n+1)&\textrm{ if
}\gamma=\gamma'+1\\
\beta + \omega^{\gamma[n]}&\textrm{ otherwise.}
\end{array}\right.\]

We define $\alpha'\rh\alpha$ whenever there is $k$ such that $\alpha' =
\alpha[k]$, or if $\alpha'+1 = \alpha$.

For instance, the fundamental sequence of $\omega$ is the sequence of
integers starting from 1. The sequence of $\ww$ is therefore
$(\omega,\omega^2,\omega^3,\dots)$. The fundamental sequence merged
with the successor relation yields for instance \[0\prec
1\prec\omega\prec\omega+1\prec\omega.2\prec\omega^2\prec\ww.\] Taking
the transitive closure of this relation gives back the original order,
so there no information loss.

\newcounter{transitive}
\setcounter{transitive}{\value{lemm}}

\begin{lemm}
\label{transitive}
The transitive closure of $\prec$ is $<$.
\end{lemm}

Moreover, the relation is \emph{crossing-free} as described below,
which is a helpful technical tool.

\newcounter{crossing-free}
\setcounter{crossing-free}{\value{lemm}}

\begin{lemm}\label{crossing-free}
If $\alpha_1<\lambda_1<\alpha_2$, $\alpha_1\prec\alpha_2$ and
$\lambda_1\prec\lambda_2$, then $\lambda_2\leq\alpha_2$.
\end{lemm}

This is the forbidden case :
\begin{displaymath}
\xymatrix{
  \alpha_1   \ar@/^1pc/[rr]
& \lambda_1\ar@/^1pc/[rr]
&  \alpha_2 
& \lambda_2
}
\end{displaymath}

\subsection{Covering graphs}

Let $\g\alpha =
\{\lambda_1\rh\lambda_2\,|\,\lambda_1,\lambda_2<\alpha\}$ be the graph
of successor and fundamental sequence relation, or \emph{covering
  graph} of the ordinal $\alpha$. For instance, a representation of
$\g\ww$ is given in Figure 1.

\begin{figure}
\label{wwgray}
\includegraphics[width=.9\textwidth]{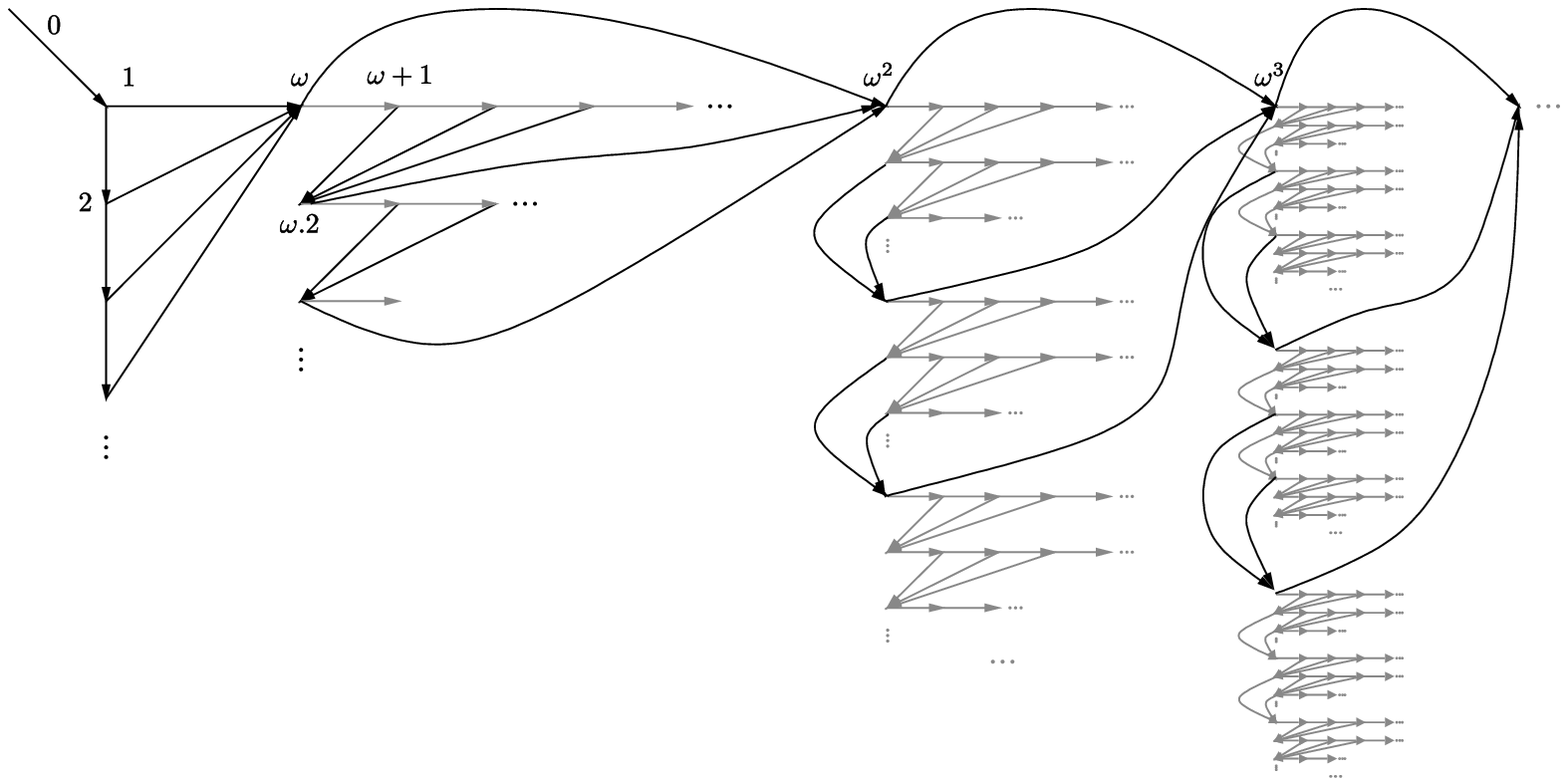}
\caption{covering graph of $\ww$.}
\end{figure}

We first remark the finite out-degree of the covering graphs. 

\newcounter{degree}
\setcounter{degree}{\value{lemm}}

\begin{lemm}
\label{degree}
For any $\omega\II (n-1)<\alpha\leq\omega\II n$ and $n>0$, the
  out-degree of $\g\alpha$ is $n$.
\end{lemm}

In the following, we refine this property to get a characterisation of
an ordinal by the degree of its vertices.  We define the \emph{degree
  word} $u(\alpha)$ of a covering graph as follows. Consider the
\emph{greatest sequence} $\sigma$ of $\g\alpha$ starting from 0,
i.e. $\sigma_0=0$ and for $k\geq0$, $\sigma_{k+1}$ is the greatest
such that $\sigma_k\rh\sigma_{k+1}$. The previous lemma ensures that
$\{\lambda\,|\,\sigma_k\prec\lambda\}$ is finite, so $\sigma_{k+1}$
exists. Such a sequence may be finite.

The degree word $u(\alpha)$ is a finite or infinite word over $[0,n]$
when $\alpha\leq \omega\II n$, and its $k^\textnormal{th}$ letter is
the out-degree of $\sigma_k$ in $\g\alpha$.

For instance, consider $u(\ww)$. Its greatest sequence is
$(0,1,\omega,\omega^2,\omega^3,\dots)$, where all have degree 2 in
$\g\ww$ except the first; so $u(\ww) = 12^\omega$. Now consider
$u(\omega^3+\omega^2)$\,: the sequence is now
\[ 0, 1, \omega, \omega^2, \omega^3, \omega^3+1,\omega^3+\omega, \omega^3+\omega+1,\dots\]
which loops into $(\dots,\omega^3+\omega.k,\omega^3+\omega.k+1,\dots)$
so $u(\omega^3+\omega^2) = 12221(21)^\omega$.

\newcounter{periodic}
\setcounter{periodic}{\value{lemm}}
\begin{lemm}
\label{periodic}
For any $\alpha\le\w\II n$, if $\alpha$ is successor then $u(\alpha)$
is a finite word of $[0,n]^*$; otherwise $u(\alpha)$ is an ultimately
periodic word of $[1,n]^\omega$.
\end{lemm}

\proofsketch If $\alpha$ is successor, then since the greatest
sequence is unbounded, the predecessor of $\alpha$ is in it and the
word is finite. Otherwise, we prove that $\alpha[k]$ is in the
greatest sequence of $\alpha$ for all finite $k$. The sequence of
degrees from $0$ to $\alpha[0]$ forms the static part of the ultimately
periodic word, whereas the sequences of degrees between $\alpha[k]$ and
$\alpha[k+1]$ are always the same.
\qed

Let $<_{lex}^n$ be the lexicographic ordering on words on $[0,n]$
based on standard order. Degree words differ for each ordinal.

\begin{lemm}
  \label{lex}
  If $\alpha<\alpha'\leq\wII n$, then $u(\alpha)<_{lex}^nu(\alpha')$.
\end{lemm}

\begin{proof}
Consider $n>0$, otherwise its degree word of $\alpha$ is the empty
word.  As before, note that the greatest sequence is unbounded, and
that $\sigma_0 = \sigma'_0 = 0$. Thus if $0<\alpha<\alpha'$ and
$\sigma'$ is the greatest sequence of $\g{\alpha'}$, there is a
smallest $n>0$ such that $\sigma_n\neq\sigma'_n$, or $\sigma_n$
doesn't exist whereas $\sigma'_n$ does. In both cases, the output
degree of $\sigma_{n-1}$ is less in $\g\alpha$ than in $\g{\alpha'}$,
so $u(\alpha)<_{lex}^nu(\alpha')$.
\qed\end{proof}

A ultimately periodic pattern can be captured by a monadic
formula. This is the goal of the the following lemma.

\begin{lemm}
\label{phi_u}
For each finite or infinite word $u$ over $[0,n]$ and a given ordinal
$\alpha$, there is a monadic formula $\phi^u$ such that
$\g\alpha\models\phi^u$ iff $u = u(\alpha)$.
\end{lemm}

\begin{proof}
The fact that the degree word is finite or ultimately periodic permits
to use a finite number of variables. We consider the ultimately
periodic case, and $u(\alpha)=uv^\omega$.

To simplify the writing, we consider the following shortcuts :
\begin{itemize}
\item $\tau(p,q)$ stands if $q$ is the greatest such that $p\rh q$;
\item if the output degree of $p$ is $k$, then $\partial_k(p)$ is true; 
\item root$(X,p)$ and end$(X,p)$ are true when $p$ is co-accessible
  (resp. accessible) from each vertex of $X$, with the entire path in
  $X$; root$(p)$ looks for a root of the whole graph;
\item inline$(X)$ checks that $X$ is a finite or infinite path;
\item size$_k(X)$ stands for $|X|=k$. 
\end{itemize}
All these notations stand for monadic formulas. For instance, the
inline$(X)$ property is true when there is a root in $X$ and each
vertex has output degree 1, and each except the root has input degree
1.

Now we may write the formula $\phi^u$. For this, we need two finite
sets $p_1\dots p_{|u|}\in U$ for the static part, $q_1\dots q_{|v|}\in
V'$ for the beginning of the periodic part and an infinite set $V$
with $V'\subseteq V$. We check that $p_1$ is the general root 0, and
$q_1$ the root of $V$, which is an infinite path. Formulas $\tau$ and
$\partial_k$ force the degree of the $uv$ part. For the periodic part,
each $q\in V$ there must be the root of a finite path $X_q\subseteq V$
of size $|v|+1$, which end has the same degree that $q$.
\qed\end{proof}

The combination of Lemmas \ref{lex} and \ref{phi_u} yields the
following theorem.

\begin{theo}
\label{th1}
For $\alpha\neq\alpha'$ smaller than $\e$, we have
$\MTh(\g\alpha)\neq\MTh(\g{\alpha'})$.
\end{theo}

As a consequence, there is no generic monadic interpretation (see next
section for definition) from an ordinal greater than $\ww$ to its
covering graph. Below this limit, there is an interpretation, because
it is possible to distinguish successive limit ordinals.

\section{The pushdown hierarchy}
\label{hierarchy}

In this section, the pushdown hierarchy will only be defined by
monadic interpretations and the treegraph operation. For other
definitions, see for instance \cite{regularpushdown}. In particular,
each level can be defined as the set of transition graphs (up to some
closure operation) of finite-state higher-order pushdown automata of
level $n$ ($n$-\emph{hopda}), hence the name. 

A major property shared by this class of graphs is the decidability of
their monadic theories. Since it is also the case for countable
ordinals \cite{shelah,buchi73}, it is natural to examine the
intersection. Here, covering graphs and ordinals are located at each
level of the hierarchy.

\subsection{Definitions}
A \emph{monadic interpretation} $I$ is a finite set $\{
\phi_a(x,y)\}_{a\in\Gamma}$ of monadic formulas with two free first
order variables. The interpretation of a graph $G\subseteq
V\times\Sigma\times V$ by $I$ is a graph $I(G)=\{p\vers{a}q
\;|\;p,q\in V \land G\vDash \phi_a(p,q)\}\subseteq V\times\Gamma\times
V$. It is helpful to have $\Gamma=\Sigma$ to allow iteration
process. The set of monadic interpretations $\mathcal I$ is closed by
composition.

A particular case of monadic interpretation is inverse rational
mapping. The alphabet $\bar\Sigma$ is used to read the arcs
backwards~: $p\vers{\bar a}q$ iff $q\vers{a}p$. An inverse rational
mapping is an interpretation such that $\phi_a(p,q)\equiv p\chemin{L_a}q$ where
$L_a$ is a regular language over $\Sigma\cup\bar\Sigma$.

For instance, the transitive closure of $R_a$ for a label $a$ is a
monadic interpretation. By Lemma \ref{transitive}, there is therefore
an immediate monadic interpretation from $\g\alpha$ to $\alpha$.  An
important corollary of Lemma \ref{th1} is that the reverse cannot
exist, or there would be a monadic formula identifying a specific
ordinal smaller than $\e$, which is contradictory to the result of
Büchi \cite[Th. 4.9]{buchi73} cited in introduction.

For a more complex illustration of a monadic interpretation, we notice
that the degree word allows the restriction from a greater ordinal.

\begin{lemm}\label{restriction}
If $\alpha<\alpha'$, there is a MSO interpretation $I$ such that
$\g\alpha = I(\g{\alpha'})$.
\end{lemm}

\begin{proof}
Following the definition, we look for an interpretation
$I=\{\psi_\prec\}$. We use again the fact that the degree word is
unique and MSO-definable. Defining the greatest sequence of $\g\alpha$
provides a MSO marking on $\g\alpha'$, which bounds the set of
vertices.  More precisely, let $\Psi^u(p)$ be an expression similar to
$\phi^u$ of the Lemma \ref{phi_u} but where the part
$\tau(p_i,p_{i+1})\land\delta_{u_i}(p_i)$ has been replaced by
$\tau_{u_i}(p_i,p_{i+1})$ meaning ``$p_{i+1}$ is the $u_i^{th}$ such
that $p_i\prec p_{i+1}$''; the same goes for the $q_j$ and for
$\tau(p_{|u|},q_1)\land\delta_{u_{|u|}}(p_{|u|})$.  Also add the
condition that $p$ is a part of the sequence~: $(\bigvee_i p=p_i)\lor
p\in V$. Then $\Psi^u(p)$ is a marking of the greatest sequence
associated to $u$. For a given $\alpha$, $I$ simply adds the condition
of co-accessibility to a vertex marked by $\Psi^{u(\alpha)}$.
\begin{eqnarray*}
\psi_\prec(p,q) &\equiv& \,p\vers{\rh}q\land\exists r\,(\Psi^{u(\alpha)}(r)\land q\vers{\rh^*}r)\\
\g\alpha &=& \{p\vers{\rh}q\,|\,p\vers{\rh}q\in \g{\alpha'}\land\exists
r\,(\Psi^{u(\alpha)}(r)\land q\vers{\rh^*}r)\}
\end{eqnarray*}
\qed\end{proof}

The \emph{treegraph} $\treegraph(G)$ of a graph $G$ is the set
$\{p\vers{a}q\}\subseteq V_G^*\times (\Sigma_G\cup\{\#\})\times V_G^*$
where $(p,q)\in V_G^*$ are sequences of vertices of $G$, and
$a\in\Sigma_G$ either if $p=wu$, $q=wv$ and $u\vers{a}v\in G$, or if
$a=\#$, $p=wu$ and $q=wuu$. One can also see the treegraph as the
fixpoint of the operation which, to each vertex which is not starting
point of an $\#$ arc, adds this arc leading to the location of this
vertex in a copy of $G$. The starting graph is called the \emph{root
  graph}.

One way to define the \emph{pushdown hierarchy} (see
\cite{carayolwohrle} for details) is as follows.
\begin{itemize}
\item $\mathcal H_0$ is the class of graphs with finite support,
\item $\mathcal H_n = \mathcal I\circ Treegraph(\mathcal H_{n-1})$.
\end{itemize}

For instance, $\mathcal H_1$ is the class of
\emph{prefix-recognizable} graphs \cite{caucal03} and further
$\mathcal H_n$ classes have been proved to correspond to an extension
of prefix-recognizability on higher-order stacks
\cite{regularpushdown}.

\subsection{Building covering graphs}

We note $p\vers{a^\bullet}q$ for the longest possible path labeled by
$a$, and $p\vers{S}q$ a shortcut for the successor relation, i.e.
\begin{eqnarray*}
p\vers{a^\bullet}q &\equiv& p\vers{a^*}q\land \neg\exists r\;(q\vers{a}r)\\
p\vers{S}q &\equiv& p\vers{\rh}q\land\neg\exists r(p\vers{\rh}r\land
r\vers{\rh^*}q).\\
\end{eqnarray*}
Now let $I=\{\phi_\rh\}$ and $M(p)$ respectively be
the interpretation and marking
\begin{eqnarray*}
\phi_\rh(p,q) &\equiv& M(p)\land M(q)\land p\chemin{\bar\rh^\bullet\#}q\lor p\chemin{\bar\#^\bullet S\#}q\lor p\chemin{\bar\#\rh\#}q\\
M(p) &\equiv& \exists r : \forall q\,(r\chemin{(\prec+\#+\bar\prec)^*}q) \land r\chemin{\rh^*\#(\bar\rh^*\#)^*}p
\end{eqnarray*}

The marking $M(p)$ allows to start anywhere on the root graph, but as
soon as a $\#$-arc has been followed, $\prec$-arcs can only be
followed backwards. We consider only goals of a $\#$-arc.

The $\phi_\prec(p,q)$ formula states the relation on these vertices,
leaving three choices~: either to follow $\prec$-arcs as long as possible
(in practice, until a copy of 0) and go down one $\#$-arc; or on the
contrary, to follow $\#$ backwards as long as possible, then take the
successor and one $\#$-arc; or just to follow one $\#$ backwards, one
$\prec$ and one $\#$.

\newcounter{phi}
\setcounter{phi}{\value{lemm}}
\begin{lemm}
\label{phi}
$\g{\omega^\alpha} = I\circ\treegraph(\g\alpha)$.
\end{lemm}

For instance consider $\g\omega$, which is an infinite path. A
representation of its treegraph is given below (plain lines for
$\rh$, dotted lines for $\#$). The circled vertices are the ones
marked by $M$ and therefore they are the only ones kept by the
interpretation $\phi$. We are allowed to go anywhere on the root
$\g\omega$ structure, but as soon as we follow $\#$ we can only go
backwards. This reflects the construction of a power of $\omega$ as a
decreasing sequence of ordinals : we may start by any, but afterwards
we only may decrease.

\begin{center}
\includegraphics[width=.8\textwidth]{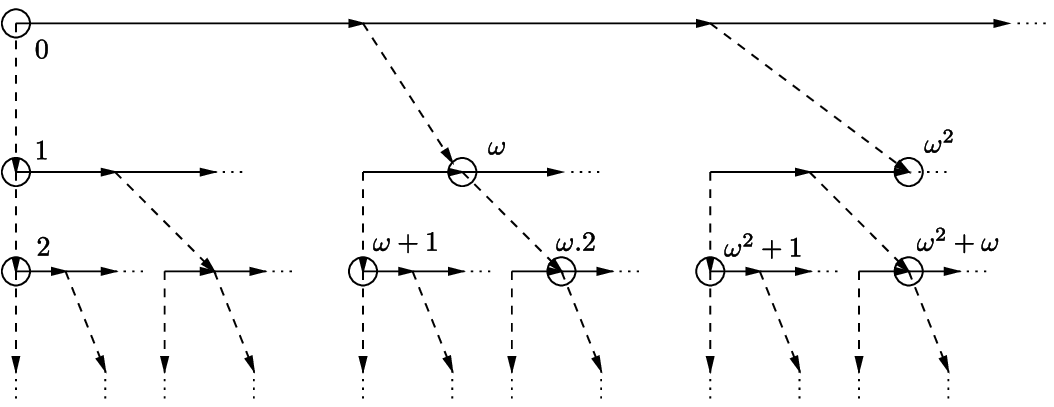}
\end{center}

\begin{lemm}
\label{ginh}
If $\alpha<\omega\II (n+1)$, then $\g\alpha\in\mathcal H_n$.
\end{lemm}

\begin{proof}
For any finite $\alpha$, $\g\alpha$ is in fact a finite path labeled
by $\rh$ and is in $\mathcal H_0$. By Lemma \ref{phi} iterated $n$
times, every $\omega^{\dots^{\omega^k}}$ with $n$ times $\omega$ and
$1<k<\omega$ is in $\mathcal H_n$.  Smaller ordinals are captured by a
restriction as in Lemma \ref{restriction}.
\qed\end{proof}

This proves the decidability of the monadic theory of the covering
graphs. By transitive closure (Lemma \ref{transitive}), ordinals are
also captured.

\begin{theo}
\label{ordinalinh}
If $\alpha<\omega\II (n+1)$, then $\alpha\in\mathcal H_n$.
\end{theo}

The decidability of the monadic theory of these ordinals is
well-known, but this result also shows that ordinals below $\e$ can be
expressed by finite objects, namely higher-order pushdown
automata. Following the steps of a well-chosen automaton (up to an
operation called the $\varepsilon$-closure) builds exactly an ordinal.
This approach is explained in Section \ref{piles}.

\section{Strictness of the hierarchy for covering graphs}
\label{strict}

In this section, we strengthen Lemma \ref{ginh} by proving that
covering graphs cannot be in any level of the hierarchy.  Let
$\exp(x,n,k)$ be a tower of exponentiation of $x$ of height $n$ with
power $k$ on the top, where $n$ and $k$ are integers.

\[\begin{array}{rcll}
\exp(x,n,k) &=& k &\textrm{ if }n=0\\
&=& x^{\exp(x,n-1,k)} & \textrm{ otherwise.}\\
\end{array}\]

In the following section, this function will be used in the cases $x=2$ and
$x=\w$.

We examine the tree $\mathcal T_n$ of trace (from the root)
$\{a^nb^{\exp(2,n,k)}\}$. It has the form below with $f(k) =
\exp(2,n,k)$. The horizontal arcs are labeled by $a$ and the vertical
arcs by $b$.

\begin{center}
\includegraphics[height=2.5cm]{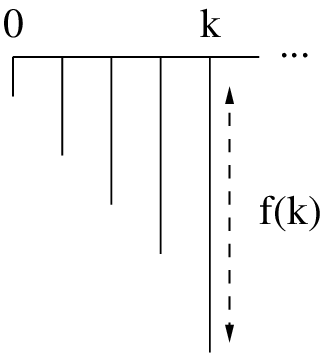}
\end{center}

For any $n$, there is such a tree which is not in the level $n$ of
the hierarchy \cite{pumpingnew}.

\begin{prop}
For $n\geq1$, $\mathcal T_{3n}\notin\graph{n}$.
\end{prop}

Finding a monadic interpretation from $\g\alpha$ to $\mathcal T_{3n}$
is therefore enough to prove $\g\alpha\notin\graph{n}$. In fact, Lemma
\ref{restriction} already states that if $\wII 3n+1\leq\alpha$, then
there is an interpretation from $\g\alpha$ to $\g{\wII 3n+1}$; so the
interpretation from $\g{\wII 3n+1}$ to $\mathcal T_{3n}$ is enough for
a whole class of ordinals. We sketch this interpretation.

Let $C^k_n$ be the set of ordinals smaller than $\exp(\w,n,k)$ where
each coefficient in RCNF is at most 1, except for the top-most power~:
\begin{itemize}
\item $[0,k-1]\in C^k_0$,
\item $0\in C^k_n$,
\item if $\gamma_0,\dots,\gamma_h$ are all distinct ordinals of $C^k_{n-1}$,
  then $\w^{\gamma_0}+\dots+\w^{\gamma_h}\in C^k_n$.
\end{itemize}

For instance, $C^3_1 = \{0,1,\w,\w+1,\w^2,\w^2+1,\w^2+\w,\w^2+\w+1\}$; 
\begin{eqnarray*}
C^2_2 &=& \{0,1,\w,\w+1,\ww,\ww+1,\ww+\w,\ww+\w+1,\\
&&\w^{\w+1},\w^{\w+1}+1,\w^{\w+1,}+\w,\w^{\w+1}+\w+1,\\
&&\w^{\w+1}+\ww,\w^{\w+1}+\ww+1,\w^{\w+1}+\ww+\w,\w^{\w+1}+\ww+\w+1\}.\\
\end{eqnarray*}

The following lemma is only a matter of cardinality of powersets.

\begin{lemm}
The cardinality of the set $C^k_n$ is $\exp(2,n,k)$.
\end{lemm}

We abusively note $\alpha+C^k_n$ for the set
$\{\alpha+\gamma\;|\;\gamma\in C^k_n\}$. The main difficulty of this
section is to define a monadic formula for this set.

\newcounter{cnk}
\setcounter{cnk}{\value{lemm}}

\begin{lemm}
\label{cnk}
For $n>0$, there is a monadic formula describing $\exp(\w,n,k)+C^k_n$
in $\g\alpha$, for $\alpha$ greater than $\exp(\w,n,k).2$.
\end{lemm}

These ordinals are easy to capture by previous tools. The following
lemma is a natural corollary of the proof of Lemma \ref{periodic},
since $\exp(\w,n,k)\prec\exp(\w,n,k+1)$.

\begin{lemm}
The greatest sequence of $\wII (n+1)$ is ultimately the sequence
$(\exp(\w,n,k))_{k\geq 1}$.
\end{lemm}

We may now state the main result of this section.

\begin{theo}
\label{th3}
If $n>0$ and $\alpha\geq \wII 3n+1$, then $\g\alpha\notin\graph{n}$.
\end{theo}

\proofsketch
If we concatenate the previous lemmas, it appears that 
\begin{itemize}
\item since the greatest sequence of $\alpha$ is interpretable from
  $\g\alpha$, we can extract the sequence $(\exp(\w,3n,k))_{k\geq 1}$
  from $\g{\wII 3n+1}$, which will be the ``horizontal path'' of
  $\mathcal T_{3n}$;
\item for each $\exp(\w,3n,k)$ we can also capture the associated set
  $\exp(\w,3n,k)+C_{3n}^k$ and arrange it in path. This yields the
  ``vertical path'' hanging from $\exp(\w,3n,k)$ and of length
  $\exp(2,3n,k)$.
\end{itemize}
Eventually, the monadic interpretation builds exactly $\mathcal T_{3n}$,
which is the expected result.
\qed

The covering graph $\g\e$ can be defined and has unbounded degree, but
has still the property of Lemma \ref{restriction}~: it can give any
smaller ordinal via monadic interpretation, which yields the following
result.

\begin{corr}
$\g\e$ does not belong to the hierarchy.
\end{corr}

From \cite{pumpingnew} we could actually extract the lower bound
$\mathcal T_{2n}\notin\graph n$. The conjecture is that $\mathcal
T_n\notin\graph n$, which would allow to locate exactly each covering
graph in the hierarchy.

The Theorem \ref{th3} does not apply to ordinal themselves, since
there we showed that there is no interpretation from ordinals to
covering graphs. Therefore, the question is still open, which leads to
Conjecture \ref{epsilon} at the end of this paper.

\section{Higher-order stack description of ordinals}
\label{piles}

The graph on the level $n$ of the hierarchy are also graphs (up to
$\varepsilon$-closure) of higher-order pushdown automata of level $n$
\cite{carayolwohrle}, i.e. automata which use nested stacks of stacks
of depth $n$. The construction by monadic interpretations and
unfolding could be translated into a pushdown automata
description. Instead of doing so, we use the equivalent notion of
prefix-recognizable relations \cite{regularpushdown} from
scratch. This notion offers a natural encoding of ordinals by their
Cantor normal form. Nonetheless, the associated proof is still heavy.

\subsection{Short presentation}

This section sketches a particular case of prefix-recognizable graphs.
For a complete description, see \cite{regularpushdown}. We only
consider 1-stacks (usual stacks) over an alphabet of size 1,
\emph{i.e.}  integers. The empty 1-stack is therefore noted 0. For
all $n>1$, a $n$-stack is a non-empty finite sequence of
$(n-1)$-stacks, noted $[a_1,\dots,a_m]_n$. The operations $Ops_1$ on a
1-stack are
\[\begin{array}{rcl}
\push_1(i) &:=& i+1,\\
\pop_1(i+1) &:=&i.\\
\end{array}\]
For $n>1$, the set $Ops_n$ of operations on a $n$-stack include
\[\begin{array}{rcll}
\zcopy_n([a_1,\dots,a_m]_n) &:=& [a_1,\dots,a_m,a_m]_n\\
\pop_n([a_1,\dots,a_m]_n) &:=& [a_1,\dots,a_{m-1}]_n\\
f([a_1,\dots,a_m]_n) &:=& [a_1,\dots,f(a_m)]_n\\
\end{array}\]
where $f$ is any operation on $k$-stacks, $k<n$. 

The 2-stack containing only 0 is noted $[\;]_2$, and the $n$-stack
containing only $[\;]_{n-1}$ is noted $[\;]_n$. Let also be an
identity operation $\id$ defined on all stacks.

The set $Ops_n$ forms a monoid with the composition
operation. Let $Reg(Ops_n)$ the closure of the finite subsets of this
monoid under union, product and iteration, \emph{i.e.}  the set of
\emph{regular expressions} on $Ops_n$. To each expression $E\in
Reg(Ops_n)$ we associate the set of $n$-stacks $S(E)= E([\;]_n)$ and
the set of relations on stacks $R(E)=\{(s,s')|s'\in E(s)\}$.

Given $E$ and a finite set $(E_a)_{a\in\Sigma}$ in $Reg(Ops_n)$, the
graph of support $S(F)$ and arcs $s\vers{a}s'$ iff $(s,s')\in R(F_a)$
is a \emph{prefix-recognizable graph of order $n$}.  General
prefix-recognizable graphs are exactly graphs of pushdown automata of
the same order.

\subsection{Towers of $\omega$}

We define the expressions $\dom$ and $\inc$ which
respectively fix the domain of the structure and the order
relation. In the following we also will need an expression $\dec$ to
perform the symmetric of $\inc$.  In one word, we want the structure
$\langle S(\dom(\alpha)),R(\dec(\alpha)),R(\inc(\alpha))\rangle$ to be isomorphic to the
structure $\langle \alpha,>,<\rangle$.

For $\omega$, we consider the set of all 1-stacks
(i.e. integers). In this case, $\dom(\w)$ is obtained by iterating
$\push_1$ on the empty stack. The other operations are also
straighforward.

\[\begin{array}{rcl}
\dom(\omega) &:=& \push_1^*\\
\inc(\omega) &:=& \push_1^+\\
\dec(\omega) &:=& \pop_1^+\\
\end{array}\]

We consider now any ordinal $\alpha$.  Let $n$ be the smallest value
such that $\dom(\alpha),\inc(\alpha)$ and $\dec(\alpha)$ are all in
$Reg(Ops_{n-1})$.

Let $\tail(\alpha):= \zcopy_n.(\id+\dec(\alpha))$. Informally, each
ordinal $\gamma<\omega^\alpha$ is either 0 or may be written as
$\gamma = \omega^{\gamma_0}+\dots+\omega^{\gamma_k}$ with
$\gamma_i<\alpha$; so we code $\gamma$ as a sequence of stacks
respectively coding $\gamma_0\dots\gamma_k$. The $\tail$ operation
takes the last stack (representing $\gamma_k$) and adds a stack coding
an ordinal $\leq\gamma_k$, so that the CNF constraint is
respected. For the relation $<$, $\inc$ either adds a decreasing
sequence (by \tail), or it first pops stacks, then increases a given
one before adding a tail.

\[\begin{array}{rcl}
\dom(\omega^\alpha) &:=& \dom(\alpha).\tail(\alpha)^*\\
\inc(\omega^\alpha) &:=& [\pop_n^*.\inc(\alpha)+\tail(\alpha)].\tail(\alpha)^*\\
\dec(\omega^\alpha) &:=& \pop_n^*.[\pop_n+\dec(\alpha).\tail(\alpha)^*]\\
\end{array}\]

We get this version of Theorem \ref{ordinalinh} restricted to towers
of $\w$.

\newcounter{cwIIn}
\setcounter{cwIIn}{\value{prop}}
\begin{theo}
\label{wIIn}
The graph of $\wIIn$ is isomorphic to the prefix-recognizable graph of
order $n$ with support $S(\dom(\wIIn))$ and one relation
$R(\inc(\wIIn))$.
\end{theo}

The proof of this proposition encodes exponentiation of $\w$, so the
case of all ordinals smaller than $\e$ can be obtained by encoding
also addition. This can be done with a greater starting alphabet and
using markers to differentiate each part of the addition.

\section{Perspectives}

We have defined covering graphs as graphs of fundamental sequence and
successor relations and shown the existence of a formula identifying a
covering graph among others, via the degree word. Then, the covering
graphs and the corresponding ordinals have been located in the
pushdown hierarchy according to the size in terms of tower of
$\omega$, in a strict way for the covering graph case.

Theorem \ref{ordinalinh} raises the question of the strictness of the
classification of ordinals in the hierarchy. Theorem \ref{th3}
naturally suggests that if $\alpha\geq\omega\II n$, then $\alpha$ does
not belong to $\mathcal H_{n-1}$, and therefore $\e$ is banned from
the hierarchy.

\begin{conj}
\label{epsilon}
$\e$ does not belong to the hierarchy.
\end{conj}

If this were proved, $\e$ would actually be a good candidate for
extending the hierarchy above the $\mathcal H_n$. Indeed, a current
field of research is to capture as many structures with decidable
monadic theory as possible. A way to do so would be to find an
operation extending those used in this paper~---~interpretation and
treegraph.

One can find definitions \cite{veblen} of a canonical fundamental
sequence for ordinals greater than $\e$ and therefore define covering
graphs outside of the hierarchy. For instance, one can take $\e[n] =
\wII (n+1)$. In this way, covering graphs may be defined for a large
number of ordinals; but we conjecture that the Theorem \ref{th1} does
not stand any more, i.e. for any definition of fundamental sequence,
there are two ordinals whose covering graphs have the same monadic
theories.

Also, the ability to differentiate covering graphs
smaller than $\e$ leads to check this robustness for more difficult
questions. One of them is selection in monadic theory, which is
negative for ordinals greater than $\ww$ \cite{selection}.

In another direction, it would be interesting to remove the
well-ordering property and to consider more general linear
orderings. The orders of $\mathbb Q$ and $\mathbb Z$ are obviously
prefix-recognizable. We would like to reach structures of more complex
orders.

\subsection*{Acknowledgments}

We would like to thank Didier Caucal and Arnaud Carayol for their
constant help, and the anonymous referees for their useful comments.

{\small
\bibliographystyle{plain}
\bibliography{bibli}
}

\begin{appendix}

\newcounter{end}
\setcounter{end}{\value{lemm}}

\section{Proofs}

\subsection{Proof of Lemma \ref{transitive}}
\setcounter{lemm}{\value{transitive}}

\begin{lemm}
The transitive closure of $\prec$ is $<$.
\end{lemm}

\begin{proof}
Let $0\leq\lambda_1\leq\lambda_2$ be two ordinals. We prove that
$\lambda_1\prec^k\lambda_2$ for some finite $k$ by induction on
$\lambda_2$.  If $\lambda_2$ is successor, consider $\lambda_2'$ such
that $\lambda_2'+1=\lambda_2$, so
$\lambda_2'\prec\lambda_2$. Otherwise, since the fundamental sequence
of $\lambda_2$ bounds all smaller ordinals, there is a smallest $n$
such that $\lambda_1\leq\lambda_2[n]\prec\lambda_2$, so let
$\lambda_2'=\lambda_2[n]$. In both cases, by induction
$\lambda_1\prec^{k'}\lambda_2'$ and thus
$\lambda_1\prec^{k'+1}\lambda_2$.\qed\end{proof}

\subsection{Proof of Lemma \ref{crossing-free}}
\setcounter{lemm}{\value{crossing-free}}

\begin{lemm}
If $\alpha_1<\lambda_1<\alpha_2$, $\alpha_1\prec\alpha_2$ and
$\lambda_1\prec\lambda_2$, then $\lambda_2\leq\alpha_2$.
\end{lemm}

\begin{proof}
We proceed by induction on $\alpha_1 = \beta+\omega^\gamma$,
$\gamma<\alpha_1$. Note that $\alpha_2>\alpha_1+1$. According to the
definition of $\prec$, there are two cases left. We suppose
$\lambda_1+1<\lambda_2$, otherwise the lemma is trivially true.

Let $\hat\beta$ denote the RCNF of $\beta$.
In the first case, $\alpha_2=\hat\beta+\omega^{\gamma+1}$. Then, in RCNF,
$\lambda_1=\hat\beta+\omega^\gamma.c_1+\hat\delta_1$ and
$\delta_1<\omega^\gamma$. Now if $\delta_1=0$, then $c_1>1$ and
$\lambda_2=\hat\beta+\omega^{\gamma+1}=\alpha_2$. If $\delta_1\neq0$, note
that the only part that changes between an ordinal and a member of its
fundamental sequence is the last term in RCNF. So $\lambda_2$ is
written $\hat\beta+\omega^\gamma.c_2+\hat\delta_2$ in RCNF with
$\delta_2<\omega^\gamma$, and therefore $\lambda_2<\alpha_2$.

In the second case, $\alpha_2=\hat\beta+\omega^{\gamma'}$ with
$\gamma\prec\gamma'$, $\gamma+1<\gamma'$. In RCNF,
$\lambda_1=\hat\beta+\omega^{\mu_1}.c_1+\hat\delta_1$ with
$\gamma\leq\mu_1<\gamma'$ and at least one of the following is true :
$\delta_1\neq0$, or $\gamma<\mu_1$, or $c_1>1$. Again, we have to deal
with several cases.

Either $\delta_1=0$ and $\gamma=\mu_1$; then $c_1>0$ and
$\lambda_2=\hat\beta+\omega^{\gamma+1}<\alpha_2$.

Or $\delta_1=0$ and $\gamma<\mu_1$; then
$\lambda_2=\hat\beta+\omega^{\mu_2}$ and $\mu_1\prec\mu_2$; this is where
the induction property is applied to get $\mu_2\leq\gamma'$, so $\lambda_2\leq\alpha_2$.

Finally, if $\delta_1\neq0$, as before
$\lambda_2=\hat\beta+\omega^{\mu_1}.c_2+\hat\delta_2<\hat\beta+\omega^{\gamma+1}<\alpha_2$.
\qed\end{proof}

\subsection{Proof of Lemma \ref{degree}}

\setcounter{lemm}{\value{degree}}
\begin{lemm}
For any $\omega\II (n-1)<\alpha\leq\omega\II n$ and $n>0$, the
  out-degree of $\g\alpha$ is $n$.
\end{lemm}

\begin{proof}
We take the cardinal of $\{\mu\,|\,\lambda\prec\mu<\wII n\}$ for an
upper bound of the output degree of $\lambda<\alpha$ in $\g\alpha$. If
$n=0$, $\lambda=0$ and $1=\wII 0$, so the set is empty.  For $n>0$,
let $\lambda = \beta+\omega^\gamma$ and $\lambda\prec\mu$, then either
$\mu=\lambda+1$ or $\mu = \beta+\omega^{\gamma'}$ with
$\gamma\prec\gamma'$. Since $\gamma'<\wII(n-1)$, by induction
$|\{\gamma'\,|\,\gamma\prec\gamma'<\wII (n-1)\}|\leq n-1$, which leads
to $|\{\mu\,|\,\lambda\prec\mu<\wII n\}| \leq n$.

For the lower bound, if $n=1$, then $\alpha\in[2,\omega]$, and
$0\prec1$ has degree 1.  For $n>1$, if $\alpha>\omega\II(n-1)$ then
\[\begin{array}{rcl}
\omega\II(n-2)&\rh&\omega\II(n-2) +1\\
&\rh&\omega^{\omega\II(n-3)+1}\\
&&\dots\\
&\rh&\omega^{\dots^{\omega+1}}\\
&\rh&\omega^{\dots^{\omega^2}}\\
&\rh&\omega\II(n-1)\\
\end{array}\]
so $\omega\II(n-2)$ has degree $n$ in $\g\alpha$.
\qed\end{proof}

\subsection{Proof of Lemma \ref{periodic}}

\setcounter{lemm}{\value{periodic}}
\begin{lemm}
For any $\alpha\le\w\II n$, if $\alpha$ is successor then $u(\alpha)$
is a finite word of $[0,n]^*$; otherwise $u(\alpha)$ is an ultimately
periodic word of $[1,n]^\omega$.
\end{lemm}

\begin{proof}
Lemma \ref{degree} ensures that the degree word is a word on the
alphabet $[1,n]$.  Since the transitive closure of $\g\alpha$ is
isomorphic to $\alpha$, the greatest sequence $\sigma$ of $\g\alpha$
is unbounded, i.e. $\forall\lambda<\alpha, \exists
n(\sigma_n\geq\lambda)$. In particular, if $\alpha=\lambda+1$, there
is $n$ such that $\sigma_n = \lambda$, and the sequence is finite. The
last element has out-degree 0.

If $\alpha$ is a limit ordinal, each $\alpha[n]$ must be in
$\sigma$. Indeed, let $m$ be such that $\sigma_m \leq \alpha[n] \leq
\sigma_{m+1}$; if the inequalities are strict, since
$\alpha[n]\rh\alpha$, by Lemma \ref{crossing-free}
$\sigma_{m+1}\geq\alpha$ which is a contradiction. So one of
$\sigma_m$ or $\sigma_{m+1}$ must be $\alpha[n]$.

We want now to prove that the pattern between the $(\alpha[n])_{n<\omega}$ is
always the same. Let $\alpha = \beta + \omega^{\gamma}$. As before, we
have two cases. If $\gamma = \gamma'+1$, then $\alpha[n] =
\beta + \omega^{\gamma'}.(n+1)$. Given $n$, there is
a path in the greatest sequence
\[\alpha[n]\prec\alpha[n]+\delta_1\prec\dots\prec\alpha[n]+\delta_h\prec\alpha[n+1]
\]
with $\delta_i<\omega^\gamma$ for each $i$, and in fact $\delta_{i+1}$
is the greatest such that $\delta_i\prec\delta_{i+1}$ and
$\delta_{i+1}\leq\omega^\gamma$. This defines the $(\delta_i)$
sequence independently of $n$. If $i$ is fixed a $n$ varies,
$\alpha[n]+\delta_i\prec \alpha[n]+x$ whenever $\delta_i<x$ and
$x\leq\omega^\gamma$, so the degree is still the same. The degree word
is therefore ultimately periodic.

In the second case, $\alpha[n]=\beta + \omega^{\gamma[n]}$ and
$\gamma[n]+1\leq\gamma[n+1]<\gamma$. So $\beta+\omega^{\gamma[n]+1}$
is in $V_{\g\alpha}$. Since
$\alpha[n]\prec\beta+\omega^{\gamma[n]+1}$, then the following element
of $\alpha[n]$ in the greatest sequence is greater than
$\beta+\omega^{\gamma[n]+1}$ and is therefore of the form
$\beta+\omega^\delta_1$ with $\gamma[n]\prec\delta_1$. In general
\[\alpha[n]=\beta + \omega^{\gamma[n]}\prec
\beta +\omega^{\delta_1}\prec \dots\prec \beta
+\omega^{\delta_h}\prec\alpha[n+1]\] are in the greatest
sequence. Then $\gamma[n],\delta_1,\dots,\gamma[n+1]$ are in the
greatest sequence of $\gamma$ and their output degrees are respectively the
same than those of
$\omega^{\gamma[n]},\omega^{\delta_1},\dots,\omega^{\gamma[n+1]}$ in $\alpha$,
minus 1. By induction, if the sequence of $\gamma$ is ultimately
periodic, so is the sequence of $\alpha$.
\qed\end{proof}

\subsection{Formulas of Lemma \ref{phi_u}}

\begin{eqnarray*}
\tau(p,q) &\equiv& p\prec q\land \forall r\;(p\prec r\Rightarrow r\prec^*q)\\
\partial_k(p) &\equiv& \exists q_1,\dots,q_k\;\left(\bigwedge_{i\neq j}p\prec q_i\land q_i\neq q_j\right)\\
\textnormal{root}(X,p) &\equiv& \forall q\in X, \forall Y\subseteq X\, (p\in Y\land\textnormal{closed}(Y)\Rightarrow q\in Y)\\
&&\textnormal{closed}(Y)\equiv \forall x,y\in Y\,((x\in X\land x\vers{}y)\Rightarrow y\in X)\\
\textnormal{size}_k(X) &\equiv&\exists q_1,\dots,q_k\;\left(\bigwedge_{i\neq j}q_i\neq q_j\land
\forall q\in X\;(\bigvee_{i}q= q_i)\right)\\
\textnormal{inline}(X) &\equiv& \exists r\in X\,(\textnormal{root}(X,r)\land \forall
p\in X\\&&[p=r\lor\exists!q\in X\,(q\rh p))\land
  \exists!q\in X\,(p\rh q)])
\end{eqnarray*}

In the $\phi_u$ formula, the $(p_i)_{i\leq|u|}$ form the static part,
and the $(q_i)_{i\leq|v|}$ the beginning of the periodic part $V$. The
last lines describe the periodicity of the degrees in $V$ with period
$|v|$.
  \begin{eqnarray*}
    \phi^u&\equiv&
    \exists p_1,\dots,p_{|u|},V,q_1,\dots,q_{|v|} \in V\,:\\
    &&\textnormal{root}(p_1)\land\left(\bigwedge_{i=1}^{|u|-1}\tau(p_i,p_{i+1})\land
    \partial_{u_i}(p_i)\right)\land \tau(p_{|u|},q_1)\land\partial_{u_{|u|}}(p_{|u|})\\
    &&\land\textnormal{root}(V,q_1)\land\left(\bigwedge_{i=1}^{|v|-1}\tau(q_i,q_{i+1})\land
    \partial_{v_i}(q_i)\right)\land \partial_{v_{|v|}}(q_{|v|})\\
    &&\land\textnormal{inline}(V)\land\forall q\in V, \exists X\subseteq V,q'\in X\,:\\
    &&\qquad\textnormal{inline}(X)\land \textnormal{size}_{|v|+1}(X) \land \textnormal{root}(X,q)\land\textnormal{end}(X,q')\\
    &&\qquad\land \left( \bigwedge_{k\leq n}\partial_k(q)\Rightarrow\partial_k(q')\right)\\
  \end{eqnarray*}

\subsection{Proof of Lemma \ref{phi}}

\setcounter{lemm}{\value{phi}}
\begin{lemm}
$\g{\omega^\alpha} = I\circ\treegraph(\g\alpha)$.
\end{lemm}

\begin{proof}
As stated in Section \ref{def}, $\omega^\alpha$ is isomorphic to the
set of decreasing sequences of ordinals smaller than $\alpha$ in
lexicographic order. Let $T = \treegraph(\g\alpha)$; the $0$ of the
root graph is still the only root, we call it $r$. Each $p\in V_T$
marked by $M$ can be mapped into a decreasing sequence. If
$r\chemin{\rh^*\#(\bar \rh^*\#)^*}p$, then there is a finite sequence
$(p_i)_{i\leq k}$ such that $r\chemin{\rh^*\#}p_0$,
$p_i\chemin{\bar\rh^*\#}p_{i+1}$ for $i<k$ and $p_k=p$. Each $p_i$ is
a copy of some $\gamma_i<\alpha$ with $\gamma_{i+1}<\gamma_i$, so
the mapping $p\mapsto (\gamma_0,\dots,\gamma_k)$ is bijective from
marked vertices of $T$ to decreasing sequences of $\alpha$.

The interpretation $\phi$ provides the relation to make this bijection
an isomorphism. Let $G=\phi\circ \treegraph(\g\alpha)$. We distinguish
the three cases of the definition of $\prec$.
\begin{itemize}
\item If $p\chemin{\bar\rh^\bullet\#}q$, then $q$ is mapped to
  $(\gamma_0,\dots,\gamma_k,0)$. This is the successor case
  $\beta_p+1=\beta_q$.
\item If $p\chemin{\bar\#^\bullet S\#}q$, then let $l$ be the smallest
  integer such that $\gamma_l =\gamma_{l+1} = \dots = \gamma_k$. Then
  $q$ is mapped to $(\gamma_0,\dots,\gamma_{l-1},\gamma_{l}+1)$. This
  corresponds to the case $\beta_p = \beta+\omega^{\gamma_l}.(k-l)\prec
  \beta+\omega^{\gamma_l+1}$.
\item If $p\chemin{\bar\#\rh\#}q$, then $q\mapsto
  (\gamma_0,\dots,\gamma_{k-1},\gamma)$ with $\gamma_k\rh\gamma$. The
  marking $M$ ensures that $q$ is mapped to a decreasing sequence. This
  is the recursive case, where $\beta_p=\beta+\omega^{\gamma_k}$,
  $\beta_p=\beta+\omega^{\gamma'_k}$ and $\gamma_k\prec\gamma'_k$.
\end{itemize}
\qed\end{proof}

\subsection{Proof of Lemma \ref{cnk}}

\setcounter{lemm}{\value{cnk}}
\begin{lemm}
For $n>0$, there is a monadic formula describing $\exp(\w,n,k)+C^k_n$ in the
covering graph of an ordinal greater than $\exp(\w,n,k).2$.
\end{lemm}

For any ordinal $\alpha$, we define a sequence $S_\alpha$. We note
$\tau(\alpha)$ the greatest $\gamma$ such that $\alpha\prec\gamma$.
\begin{itemize}
\item $\alpha\in S_\alpha$, $\alpha+1\in S_\alpha$,
\item if $\lambda\in S_\alpha$ and $\alpha<\lambda\prec\gamma$, then
  $\gamma\in S_\alpha$ unless $\exists \lambda'\leq\lambda$ such that
  and $\lambda'\in S_\alpha$ and $\lambda'\prec\tau(\gamma)$.
\end{itemize}

It is easy to express $S_\alpha$ with a monadic formula. It happens to
be the requested set.

\setcounter{lemm}{\value{end}}
\begin{lemm}
The set $S_{\exp(\w,n,k)}$ is $\exp(\w,n,k)+C^k_n$ in the covering
graph of an ordinal greater than $\exp(\w,n,k).2$.
\end{lemm}

\begin{proof}
Let $\alpha=\exp(\w,n,k)$. First of all, $\tau(\alpha.2) =
\w^{\exp(\w,n-1,k)+1}$ and $\alpha \prec \w^{\exp(\w,n-1,k)+1}$ so
$\tau(\alpha.2)\notin S_\alpha$. By Lemma \ref{crossing-free}, any
path from $\alpha$ to an ordinal of $[\alpha.2,\w^{\exp(\w,n-1,k)+1}[$
    goes through $\alpha.2$, and paths to ordinals of
    $[\w^{\exp(\w,n-1,k)+1},\exp(\w,n,k+1)[$ go through a successor of
        $\alpha$ which is not in $S_\alpha$, so
        $S_\alpha\cap[\alpha.2,\exp(\w,n,k+1)]=\emptyset$.

\begin{displaymath}
\xymatrix{
  \alpha   \ar@/^1pc/[rr]
\ar@/^1.5pc/[rrr]
&  \alpha.2 \ar[r]
&  \w^{\exp(\w,n-1,k)+1}
& \dots
}
\end{displaymath}

Let $\lambda\in[\alpha,\alpha.2[$, $\lambda = \hat\beta +
    \w^\gamma.c+\hat\eta$ in RCNF with $c>1$ (as in Lemma
    \ref{crossing-free}, we use the notation $\hat\beta$ to note the
    RCNF of $\beta$). By Lemma \ref{crossing-free} again, any path
    from $\alpha$ to $\alpha+\lambda$ goes through
\[\begin{array}{rrcl}
&\lambda'&=&\hat\beta + \w^\gamma\\
\textrm{and }&\lambda''&=&\hat\beta + \w^\gamma.c\\
\end{array}\]

\begin{displaymath}
\xymatrix{
  \alpha 
& \lambda' \ar@/^1.5pc/[rrr]
& \lambda'' \ar@/^1pc/[rr]
& \lambda
& \beta + \w^{\gamma+1}
& \alpha.2
}
\end{displaymath}

But then $\lambda'\prec\beta+\w^{\gamma+1} = \tau(\lambda'')$ when $c>1$,
so $\lambda''\notin S_\alpha$.

Recursively, we suppose that any path from $\exp(\w,n-1,k)$ to
$\gamma$ with $\exp(\w,n-1,k)\leq\gamma<\exp(\w,n-1,k).2$ goes through
$\gamma'$ and $\gamma''$, with $\gamma'\prec\tau(\gamma'')$. Then if
$\lambda = \hat\beta+\w^\gamma+\hat\eta$, define
\[\begin{array}{rrcl}
&\lambda'&=&\hat\beta + \w^{\gamma'}\\
\textrm{and }&\lambda''&=&\hat\beta + \w^{\gamma''}\\
\end{array}\]
which propagate the property to level $n$. All this proves that if
$\lambda = \alpha+\w^{\gamma_0}+\dots+\w^{\gamma_j}$ in CNF, then all
$\gamma_i$ are distinct and are in $C^k_{n-1}$. Therefore
$S_\alpha\subseteq\alpha+C^k_n$.

For the other side, let $\lambda\in\alpha+C^k_n$. If $\lambda=\alpha$
the case is done, otherwise
\[\lambda=\alpha+\omega^{\gamma_0}+\dots+\omega^{\gamma_h}\]
with each $\gamma_i\in C^k_{n-1}$.  

We have to prove that $\exists\lambda'\prec\lambda$ in $C^k_n$. By
induction, for $\gamma_h>0$, $\exists\gamma'\prec\gamma$ in
$C^k_{n-1}$, so
$\lambda'=\alpha+\omega^{\gamma_0}+\dots+\omega^{\gamma'}$ answers to
the question (since the $\gamma_i$ are decreasing, the ``distinct''
constraint is respected). If $\gamma_h=0$, then we take
$\lambda'=\alpha+\omega^{\gamma_0}+\dots+\omega^{\gamma_{h-1}}$.

Now
$\tau(\lambda)=\alpha+\omega^{\gamma_0}+\dots+\omega^{\tau(\gamma_h)}$.
If $\lambda'\in S_\alpha$ is such that $\lambda'\prec\tau(\lambda)$,
then
$\lambda'=\alpha+\omega^{\gamma_0}+\dots+\w^{\gamma_{h-1}}+\omega^{\gamma}$
for some $\gamma\in\gamma_h\cap C^k_{n-1}$, but then by induction we
never have $\gamma\prec\tau(\gamma_h)$, which is a contradiction.
\end{proof}

\subsection{Proof of Proposition \ref{wIIn}}
\label{proofwIIn}

\setcounter{prop}{16}
\begin{prop}
The graph of $\wIIn$ is isomorphic to the prefix-recognizable graph of
order $n$ with support $S(\dom(\wIIn))$ and one relation
$R(\inc(\wIIn))$.
\end{prop}

\proof The theorem is easy for $n=1$. Vertices of $\w$ are precisely
all 1-stacks; $R(\inc(\w))$ is the successor relation, while
$R(\dec(\w))$ is the symmetric relation.

We suppose now that $n>1$, and that there exist $\dom(\alpha),
\inc(\alpha),\dec(\alpha)$ operations in $Reg(Ops_{n-1})$ such that
$\langle S(\dom(\alpha)),R(\dec(\alpha)),R(\inc(\alpha))\rangle$ is
isomorphic to $\langle \alpha,>,<\rangle$.  We also suppose that
$\dec(\alpha)(S(\dom(\alpha)))\subseteq S(\dom(\alpha))$. For any
$\gamma<\alpha$, we note $s_\gamma$ the corresponding $(n-1)$-stack.

Note that if $k<n$, all operations on $k$-stacks are valid on
$n$-stacks. So if $f\in Reg(Ops_k)$ is and operation and $s,s'$ are two
$k$-stacks such that $(s,s')\in R(f)$ , and if $p,p'$ are the same
$n$-stack except for the top-most $k$-stack which is respectively $s$
and $s'$, then $(p,p')\in R(f)$.

Let $S=\dom(\w^\alpha)$ and let $p\in S$ be a non-empty finite
sequence of $(n-1)$-stacks, so
$p=[s_{\gamma_0},\dots,s_{\gamma_k}]$. In the definition of
$\dom(\w^\alpha)$, there is no $\pop_n$ operation, and by the
induction property and the above remark,
$s_{\gamma_0},\dots,s_{\gamma_k}$ are all in $S(\dom(\alpha))$. By
hypothesis on $\dec(\alpha)$, we also have $s_{\gamma_0}\geq\dots\geq
s_{\gamma_k})$. As a consequence, the mapping
\[
p = [s_{\gamma_0},\dots,s_{\gamma_k}]\mapsto \lambda = \w^{\gamma_0}+\dots+\w^{\gamma_k}
\]
is well defined and is injective. In fact, it is a bijection between
$S$ and $[1,\w^\alpha[$; omitting 0 is not a problem for infinite
    ordinals. We therefore note $p_\lambda$ the $n$-stack associated
    to $\lambda$.

Now if let $0<\lambda<\lambda'<\alpha$ be two ordinals, with
$\lambda=\omega^{\gamma_0}+\dots+\omega^{\gamma_k}$ in CNF. Then
\[\begin{array}{rrcll}
\textrm{either }&\lambda'&=&\omega^{\gamma_0}+\dots+\omega^{\gamma_k}+\dots+\omega^{\gamma_{k'}}&\textrm{ with }k<k',\\
\textrm{or }
&\lambda'&=&\omega^{\gamma_0}+\dots+\omega^{\gamma_i}+\dots+\omega^{\gamma'_{i+1}}+\dots+\omega^{\gamma_{k'}}&\textrm{ for some }i<k,\\
\end{array}\]
with $\gamma_{i+1}<\gamma'_{i+1}$. For the first case, the use of
$\tail(\alpha)$ on $p_\lambda$ has already been discussed, so
$(p_\lambda,p_{\lambda'})\in R(\tail(\alpha)^+)$. In the second case,
$\pop_n^{(k-i-1)}(p_\lambda) = [s_{\gamma_0},\dots,s_{\gamma_{i+1}}]$
and, by induction, 
\[([s_{\gamma_0},\dots,s_{\gamma_{i+1}}], [s_{\gamma_0},\dots,s_{\gamma'_{i+1}}])\in R(\inc(\alpha)).\]
Again, the $\tail$ operation is used. The converse~--- if
$(p_\lambda,p_{\lambda'})\in S^2\cap R(\inc(\w^\alpha))$ then $\lambda<\lambda'$~--- is
straightforward.  So $\langle
S,R(\inc(\omega^\alpha))\rangle$ is indeed
isomorphic to $\langle\alpha,<\rangle$.

The $\dec$ operation is similar. In the first case,
$\pop_n^{(k'-k)}(p_{\lambda'}) = p_{\lambda}$ with $k'-k\geq1$. In the
second case, $\pop_n^{(k'-i-1)}(p_\lambda') =
  [s_{\gamma_0},\dots,s_{\gamma'_{i+1}}]$ and
\[([s_{\gamma_0},\dots,s_{\gamma'_{i+1}}], [s_{\gamma_0},\dots,s_{\gamma_{i+1}}])\in R(\dec(\alpha)).\]
The converse is direct as well, and proves in the same time the last
needed induction property~: $\dec(\w^\alpha)(S)\subseteq S$. Note that
this was not true with \inc~: $\inc(\w^\alpha)(S)\not\subseteq S$,
because we could lose the decreasing constraint of the CNF.

Finally $\langle
S,R(\inc(\omega^\alpha),R(\dec(\omega^\alpha))\rangle$ is isomorphic
to $\langle\alpha,<,>\rangle$ and the induction properties are
fulfilled. 
\qed

\end{appendix}

\end{document}